\documentclass[letterpaper,12pt]{article}
\pdfoutput=1
\usepackage{jheppub}
\usepackage{epsfig}
\usepackage{amsmath}
\usepackage{subfigure}
\usepackage{float}
\usepackage{tikz}
\usepackage[normalem]{ulem}
\usetikzlibrary{snakes}
\usetikzlibrary{patterns}
\usetikzlibrary{decorations.pathmorphing}
\usetikzlibrary{decorations.markings}
\tikzstyle{photon}[1]=[decorate,decoration={snake,amplitude=#1}]
\tikzstyle{boson}[1]=[decorate]
\tikzstyle{gluon}[1]=[decorate,decoration={coil,aspect=1,amplitude=1.5mm}]
\tikzset{>=stealth}

\DeclareSymbolFont{AMSa}{U}{msa}{m}{n}
\DeclareSymbolFont{AMSb}{U}{msb}{m}{n}
\let\Box\relax
\DeclareMathSymbol{\Box}{\mathord}{AMSa}{"03}

\newcommand{\be}{\begin{equation}}
\newcommand{\ee}{\end{equation}}
\newcommand{\bea}{\begin{eqnarray}}
\newcommand{\eea}{\end{eqnarray}}

\newcommand{\f}{\frac}
\newcommand{\h}{\hspace*{1mm}}

\newcommand{\ds}{\displaystyle}

\newcommand{\Hd}{{\mathcal{H}}}
\newcommand{\Ld}{{\mathcal{L}}}
\newcommand{\Sd}{{\mathcal{S}}}

\newcommand{\Rd}{{\mathcal{R}}}
\newcommand{\Md}{{\mathcal{M}}}
\newcommand{\ra}{\rightarrow}

\title{Vector field instability and the primordial tensor spectrum}

\author{Stefan Eccles,}
\author{Willy Fischler,}
\author{Dustin Lorshbough}
\author{and Benjamin A. Stephens}
\affiliation{Department of Physics and Texas Cosmology Center\\ The University of Texas at Austin,
TX 78712.}
\emailAdd{stefan.eccles@utexas.edu}
\emailAdd{fischler@zippy.ph.utexas.edu}
\emailAdd{lorsh@utexas.edu}
\emailAdd{benstephens@utexas.edu}

\abstract{It has recently been shown that the presence of a spectator pseudoscalar field, coupled to photons through a Chern-Simons term, can amplify the primordial tensor spectrum without observationally disrupting the primordial scalar spectrum.  The amplification occurs due to an instability that develops for the vector fields.  We consider the extension of previous studies to account for the contribution arising from an inhomogeneous vector background which is generated prior to the onset of inflation.  We find that there may be contributions in which net momentum is transferred between the inhomogeneous vector background and the gravitons, which would give rise to a signature different than in the absence of the semiclassical corrections.  We discuss the properties the classical vector field form must have in order for these signatures to leave observable imprints, though we were unable to construct a model for generating such a vector field.}

\begin{document}
\maketitle
\flushbottom

\section{Introduction}
One of the primary tests of the theory of inflation is the detection of B-mode polarization in the cosmic microwave background (CMB) which encodes the amplitude of primordial tensor fluctuations \cite{Baumann:2008aq}.  In the simplest models of inflation, such a measurement is expected to tell us the scale of inflation $V(\phi)$.  However, the presence of additional pseudoscalar spectator fields during inflation could complicate the relationship between the tensor to scalar ratio $r$ and scale of inflation $V(\phi)$ as shown by Barnaby et al. \cite{Barnaby:2012xt}.

Carroll, Field and Jackiw discovered that in the presence of a Lorentz violating Chern-Simons term, vector fields develop instabilities \cite{Carroll:1989vb}.\footnote{For a review on how vector fields may be relevant for inflation in different contexts, see \cite{Maleknejad:2012fw} and references therein.}  This instability generically results in a spatially inhomogeneous amplification of the vector field.  When the source of the Chern-Simons term Lorentz violation is a homogeneous pseudoscalar field which is slowly rolling\footnote{The case of an inhomogeneous pseudoscalar slowly rolling will also result in a vector field instability, but we restrict our attention to the homogeneous case to make the problem more tractable.  In order to be slowly rolling the pseudoscalar potential must satisfy potential flatness conditions familiar from inflationary theory, $M_PV'\ll V$ and $M_P^2V''\ll V$.}, the instability remarkably enhances the primordial tensor spectrum as discussed by Anber and Sorbo \cite{Anber:2006xt}.

Previous studies have considered a vector field background that is either spatially homogeneous \cite{Bartolo:2014hwa} or vanishing \cite{Barnaby:2012xt}.  However, generically the instability will exponentially amplify any small deviation away from homogeneity. This may prompt the question whether there could be observable effects due to an initial inhomogeneous classical vector field configuration.  In this work, we illustrate what type of functional form the classical vector field must have to leave observable imprints.  We were unable to construct a model which gives rise to a classical vector field with the appropriate properties.

The paper is organized as follows: we review the vector field instability in flat space \cite{Carroll:1989vb} and de Sitter space \cite{Anber:2006xt} in section \ref{sec:Instability}.  In section \ref{sec:Quantum} we review contribution of quantum loops to the tensor two point function.  In section \ref{sec:Classical} we discuss the properties the classical inhomogeneous vector background must have to leave observable imprints.  We conclude in section \ref{sec:Conclusion}.
\section{Vector field instability}\label{sec:Instability}
\subsection{Vector field instability in flat space}
In this section we review the flat space vector instability discovered by Carroll, Field and Jackiw.  They studied the bounds that arise on violations of Lorentz invariance and parity by considering the addition of a Chern-Simons term to the usual Maxwell theory in the current universe \cite{Carroll:1989vb},
\begin{equation}\label{eq:CFJ_action}
\mathcal{L}=-\f{1}{4}F^2-\frac{1}{2}p_\alpha A_\beta \tilde{F}^{\alpha \beta}.
\end{equation}
We use the notation that $F_{\alpha\beta}=\left(\partial_\alpha A_\beta-\partial_\beta A_\alpha\right)$ is the usual Maxwell tensor, $\tilde{F}^{\alpha\beta}=\f{1}{2\sqrt{-g}}\epsilon^{\alpha\beta\mu\nu}F_{\mu\nu}$ is the dual to the Maxwell tensor and the Levi-Civita symbol is always the flat space one with metric factors written explicitly.  The four-vector $p_\alpha$ parameterizes the breaking of Lorentz invariance.  

The origin of the four-vector $p_\alpha$ is constrained by imposing gauge invariance, i.e. that $\Delta \Ld=0$ when $\Delta A_\mu=\partial_\mu\psi$, on the Chern-Simons term
\begin{equation}\label{eq:CFJ_gauge_inv}
\Delta \Ld_{\text{CS}}=\f{1}{4}\psi \tilde{F}^{\beta\alpha}(\nabla_\alpha p_\beta-\nabla_\beta p_\alpha)=0.
\end{equation}
Since we ultimately want to discuss this phenomenon in curved spacetimes, we use covariant derivatives.  In order to satisfy gauge invariance for general $\psi$, the difference of covariant derivatives must vanish.  We immediately see that this is satisfied if either $p_\alpha$ is a constant or the derivative of a (pseudo)scalar field, $p_\alpha=\partial_\alpha\chi$, which is what we will consider in the next section.

The vector field exhibits an instability shown \cite{Carroll:1989vb} by the dispersion relation for the field modes (for the sake of clarity we set $p_\alpha=p_0$),
\begin{equation}\label{eq:CFJ_dispersion}
\omega_k^2=k^2\left(1- \lambda\f{p_0}{k}\right).
\end{equation}
The $\lambda=+1$ polarization contains modes $k<p_0$ which experience runaway instability.

\subsection{Vector field instability during inflation}
The study of this instability was generalized by Anber and Sorbo \cite{Anber:2006xt} to the case of de Sitter evolution with a pseudoscalar field present,
\begin{equation}\label{eq:dS_general}
\Ld\supset a^3\left[-\f{1}{2}(\partial\chi)^2-V(\chi)-\f{1}{4}F^2-\f{\chi}{4f} F\tilde{F}\right].
\end{equation}
For the problem to be tractable we take the $\chi$ field to be homogeneous.  This simplification allows us to write a simplified action through integrating by parts
\begin{equation}\label{eq:dS_homo}
\Ld\supset a^3\left[\f{1}{2}\dot{\chi}^2 -V(\chi)+\f{\dot{\chi}}{2fa^3}\epsilon^{i j k} A_i \partial_j A_k\right].
\end{equation}

We choose initial conditions such that the homogeneous pseudoscalar field is slowly rolling and such that the vector backreaction onto the pseudoscalar evolution is small\footnote{We will find the backreaction bounds on the model parameters explicitly in the next section.}.  After Fourier expanding the vector field, we find the equations of motion to simplify to
\begin{equation}\label{eq:chi_eom}
\ddot{\chi}+3H\dot{\chi}+V'=0,
\end{equation}
\begin{equation}\label{eq:vector_eom}
\ddot{A}_q^\lambda+H\dot{A}_q^\lambda+\f{q^2}{a^2}\left(1-\lambda\f{2\xi}{(q/aH)}\right)A_q^\lambda=0,\h\h\h\xi=\f{\dot{\chi}}{2Hf}.
\end{equation}
The parity violation in the dispersion relation is inherited from the Chern-Simons term and may lead to unique observational signatures as emphasized by \cite{Sorbo:2011rz}.  We take $\dot{\chi}>0$ for definitiveness, then the $\lambda=+1$ polarization experiences an instability for values of $(q/aH)<2\xi$.

The expansion of the universe differentiates this case from the flat space study of \cite{Carroll:1989vb} in two important ways.  The first is that the scale factor introduces time dependence into the criteria for which modes will undergo exponential growth.  Therefore modes will continually redshift into the instability regime as inflation proceeds as long as the pseudoscalar field remains homogeneous and slowly rolling.  Furthermore, the overall scale factor suppression of the vector field frequency causes the vector field modes to eventually freeze out after some period of growth.

Previous studies have computed the contribution arising from quantum fluctuations, which we will present in the next section.  In section \ref{sec:Classical} we will discuss requirements for allowing the classical initial inhomogeneous background to have observable effects.

\section{Review of the quantum contribution}\label{sec:Quantum}
We study the inflationary model first considered by Barnaby et. al. \cite{Barnaby:2012xt} in which the field content from the previous section are spectator fields to the inflaton $\phi$,
\begin{equation}\label{eq:action}
S=\int d^4x\,\sqrt{-g}\left[{\mathop{\underbrace{\f{M_p^2}{2}R-\f{1}{2}(\partial\phi)^2-V(\phi)}_{\text{Usual Inflationary Theory}}}}\mathop{\underbrace{-\f{1}{2}(\partial\chi)^2-V(\chi)-\f{1}{4}F^2-\f{1}{4f}\chi F\tilde{F}}_{\text{Spectator Axionic Sector}}}\right].
\end{equation}

The condition that the kinetic energy of the pseudoscalar field be subdominant to that of the inflaton ensures that the comoving curvature perturbation is dominated by the inflaton field contribution.  For two homogeneous fields the scalar curvature perturbation and its evolution are given as \cite{GarciaBellido:1995fz,Taruya:1997iv,Finelli:2000ya}
\begin{equation}\label{eq:R_2fld}
\Rd=\f{H}{\left(\dot{\phi}^2+\dot{\chi}^2\right)}\left(\dot{\phi}\h Q_\phi+\dot{\chi}\h Q_\chi\right)\rightarrow \f{H}{\dot{\phi}}Q_\phi\text{ if }\dot{\chi}\ll\dot{\phi},
\end{equation}
\begin{equation}\label{eq:Rdot_2fld}
\dot{\Rd}=\left(\begin{array}{c}\text{Spatial}\\\text{Gradient}\end{array}\right)+\f{H}{2}\left(\f{Q_\phi}{\dot{\phi}}-\f{Q_\chi}{\dot{\chi}}\right)\f{d}{dt}\left(\f{\dot{\phi}^2-\dot{\chi}^2}{\dot{\phi}^2+\dot{\chi}^2}\right).
\end{equation}
In the $\dot{\chi}\ll\dot{\phi}$ limit we recover the single field comoving curvature perturbation behavior expected for single field inflation.  We have used $Q_{f}=H^{-1}\dot{f}\Psi+\delta f$ to denote the field perturbations in the spatially flat gauge and $\Psi$ as the longitudinal gauge spatial metric perturbation.  Typically, the additional degree of freedom is expressed as an isocurvature perturbation for two fields with either the following normalization \cite{Gordon:2000hv,Amendola:2001ni,Malik:2008im},
\begin{equation}\label{eq:S_2fld}
\Sd=\f{H}{\left(\dot{\phi}^2+\dot{\chi}^2\right)}\left(-\dot{\chi}\h \delta\phi+\dot{\phi}\h \delta\chi\right)\rightarrow \f{H}{\dot{\phi}}\delta\chi\text{ if }\dot{\chi}\ll\dot{\phi},
\end{equation}
or an alternative normalization,
\begin{equation}
S_{\chi\phi}=\f{\left(\dot{\phi}^2+\dot{\chi}^2\right)}{\dot{\phi}\dot{\chi}}\Sd=H\left(\f{\delta \chi}{\dot{\chi}}-\f{\delta \phi}{\dot{\phi}}\right)\ra\f{H}{\dot{\chi}}\delta\chi\text{ if }\dot{\chi}\ll\dot{\phi}.
\end{equation}

Recently, Ferreira and Sloth \cite{Ferreira:2014zia} have shown that when rewriting the action in terms of the isocurvature perturbation there exists a vertex which couples $\Rd$ and $F\tilde{F}$.  In particular, notice that for the conventions we have introduced we may rewrite the pseudoscalar fluctuation in terms of the isocurvature perturbation as
\begin{equation}
Q_\chi=\f{\dot{\phi}}{H}\left(\Sd+\f{\dot{\chi}}{\dot{\phi}}\Rd\right)=\f{\dot{\phi}}{H}\f{\dot{\chi}}{\dot{\phi}}\left(\f{\Sd_{\chi\phi}}{\left(1+\dot{\chi}^2/\dot{\phi}^2\right)}+\Rd\right).
\end{equation}
It was shown that the existence of the $\Rd F\tilde{F}$ coupling places a strong bound of the amount of vector field amplification, even if the pseudoscalar is only coupled gravitationally to the inflaton field.  The case of allowing the pseudoscalar to roll for a small number of efolds was discussed and recently studied in greater detail by \cite{Namba:2015gja}.

\begin{figure}
\centering
\begin{tikzpicture}
\draw[boson] (0,0) -- (1,0);
\draw[photon] (1,0) -- (2,0);
\draw[photon] (1,0) -- (1,1) node{$\times$};
\draw[photon] (2,0) -- (2,1) node{$\times$};
\draw[boson] (2,0) -- (3,0);
\draw[fill=black] (1,0) circle (2pt);
\draw[fill=black] (2,0) circle (2pt);
\draw (1.5,-0.45) node{\bf ($\Md_2^\Rd$)};
\end{tikzpicture}~~~~~~~~~\begin{tikzpicture}
\draw[boson] (0,0) -- (1,0) ;
\draw[boson] (1,0) --  (2,0);
\draw[photon] (1,0) -- (0.5,1) node{$\times$};
\draw[photon] (1,0) -- (1.5,1) node{$\times$};
\draw[fill=black] (1,0) circle (2pt);
\draw (1,-0.45) node{\bf ($\Md_1^\Rd$)};
\end{tikzpicture}~~~~~~~~~\begin{tikzpicture}
\draw[boson] (0,0) -- (1,0);
\draw[photon] (1.5,0) circle (0.5);
\draw[boson] (2,0) -- (3,0);
\draw[fill=black] (1,0) circle (2pt);
\draw[fill=black] (2,0) circle (2pt);
\draw (1.5,-0.75) node{\bf ($\Md_2^\Rd$)};
\end{tikzpicture}~~~~~~~~~\begin{tikzpicture}
\draw[boson] (0,0) -- (1,0) ;
\draw[boson] (1,0) --  (2,0);
\draw[photon] (1,0.5) circle (0.5);
\draw[fill=black] (1,0) circle (2pt);
\draw (1,-0.375) node{\bf ($\Md_1^\Rd$)};
\end{tikzpicture}
\begin{tikzpicture}
\draw[gluon] (0,0) -- (1,0);
\draw[photon] (1,0) -- (2,0);
\draw[photon] (1,0) -- (1,1) node{$\times$};
\draw[photon] (2,0) -- (2,1) node{$\times$};
\draw[gluon] (2,0) -- (3,0);
\draw[fill=black] (1,0) circle (2pt);
\draw[fill=black] (2,0) circle (2pt);
\draw (1.5,-0.45) node{\bf ($\Md_2^h$)};
\end{tikzpicture}~~~~~~~~~\begin{tikzpicture}
\draw[gluon] (0,0) -- (1,0) ;
\draw[gluon] (1,0) --  (2,0);
\draw[photon] (1,0) -- (0.5,1) node{$\times$};
\draw[photon] (1,0) -- (1.5,1) node{$\times$};
\draw[fill=black] (1,0) circle (2pt);
\draw (1,-0.45) node{\bf ($\Md_1^h$)};
\end{tikzpicture}~~~~~~~~~\begin{tikzpicture}
\draw[gluon] (0,0) -- (1,0);
\draw[photon] (1.5,0) circle (0.5);
\draw[gluon] (2,0) -- (3,0);
\draw[fill=black] (1,0) circle (2pt);
\draw[fill=black] (2,0) circle (2pt);
\draw (1.5,-0.75) node{\bf ($\Md_2^h$)};
\end{tikzpicture}~~~~~~~~~\begin{tikzpicture}
\draw[gluon] (0,0) -- (1,0) ;
\draw[gluon] (1,0) --  (2,0);
\draw[photon] (1,0.5) circle (0.5);
\draw[fill=black] (1,0) circle (2pt);
\draw (1,-0.375) node{\bf ($\Md_1^h$)};
\end{tikzpicture}
\caption{Semiclassical and quantum loop diagrams generated by first and second order interaction terms.  The two vertex diagram ($\Md_2$) contains four enhanced vector fields whereas the one vertex diagram ($\Md_1$) contains two, therefore the two vertex diagram is the dominant contribution to the primordial spectra.}
\label{fig:diagrams}
\end{figure}
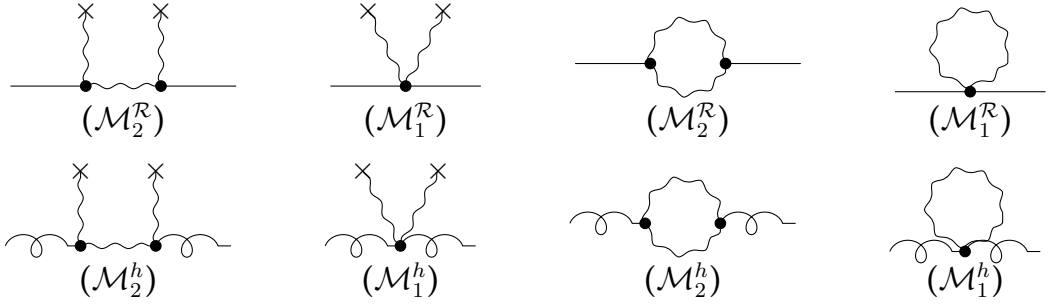

We will now calculate the enhanced primordial tensor spectrum that results from the vector field instability.  We take the usual vacuum mode function tensor fluctuations
\begin{equation}
h_k^s(t)=\frac{H}{M_Pk^{3/2}}\left(1+\frac{ik}{aH}\right)e^{-ik/aH}.
\end{equation}
We use the ``in-in'' formalism with the master equation (for any operator W(t)) \cite{Weinberg:2005vy,Adshead:2009cb,Weinberg:2010wq}
\begin{equation}\label{eq:inin_Master}
\langle W(t)\rangle=\left\langle\left(\bar{T}e^{i\int_{-\infty}^t H_{\text{int}}(t')dt'}\right) W(t)\left(Te^{-i\int_{-\infty}^t H_{\text{int}}(t')dt'}\right)\right\rangle.
\end{equation}
The Feynman rules may be derived from the relevant first and second order interaction Hamiltonians
\begin{equation}
\begin{array}{ll}
\Hd_{\text{h,int}}^{(1)}\supset&-\f{1}{2}a^5g^{\alpha\rho}g^{\mu\sigma}\left(g^{\beta\nu}+2\delta^{\beta0}\delta^{\nu0}\right)h_{\rho\sigma}F_{\mu\nu}F_{\alpha\beta},\\
\Hd_{\text{h,int}}^{(2)}\supset&-\f{1}{16}a^7\left[4g^{\alpha\rho}g^{\mu\sigma}g^{\beta\omega}g^{\nu\theta}h_{\rho\sigma}h_{\omega\theta}+8g^{\alpha\rho}g^{\mu\sigma}g^{\omega\theta}\left(g^{\beta\nu}+2\delta^{\beta0}\delta^{\nu0}\right)h_{\rho\omega}h_{\sigma\theta}\right.\\
&\left.-g^{\alpha\mu}g^{\rho\sigma}g^{\omega\theta}\left(g^{\beta\nu}+4\delta^{\beta0}\delta^{\nu0}\right)h_{\rho\theta}h_{\omega\sigma}\right]F_{\mu\nu}F_{\alpha\beta}.
\end{array}
\end{equation}
The perturbative expansion of (\ref{eq:inin_Master}) leads to two different diagram topologies (see figure \ref{fig:diagrams}).  There is a single vertex diagram ($\Md_1$) arising from one insertion of the second order interaction Hamiltonian and a two vertex diagram ($\Md_2$) arising from two insertions of the first order interaction Hamiltonian.

The quantum loop calculation of the two vertex diagram with which we will compare has been treated in detail using an alternate formalism \cite{Barnaby:2012xt,Anber:2012du,Barnaby:2011vw,Sorbo:2011rz,Bartolo:2014hwa}, while the $\Md_1$ diagram has usually been assumed negligible for large $\xi$.  This is reasonable since $\Md_1$ arises from a single interaction term of the form $(h h A A)$ while $\Md_2$ contains the product $(h A A)^2$.  The number of vector modes present in each case sets the relative size of the diagrams: $\Md_1$ $\propto e^{2\pi\xi}$ while $\Md_2$ $\propto e^{4\pi\xi}$.  We find (in the helicity basis) the following contribution from each diagram:
\begin{equation}\label{eq:QuantumPS}
\begin{array}{ccl}
\ds{\langle\hat{h}^+(\vec{k}_1,t)\hat{h}^+(\vec{k}_2,t)\rangle_{\Md_2}^{\text{quant}}}&=&\ds{(2\pi)^3\delta^3\left(\vec{k}_1+\vec{k}_2\right)\times\left(8.6\times10^{-7}\right)\f{H^4}{M_P^4k^3}\f{e^{4\pi\xi}}{\xi^6}},\\
\ds{\langle\hat{h}^-(\vec{k}_1,t)\hat{h}^-(\vec{k}_2,t)\rangle_{\Md_2}^{\text{quant}}}&=&\ds{(2\pi)^3\delta^3\left(\vec{k}_1+\vec{k}_2\right)\times\left(1.8\times10^{-9}\right)\f{H^4}{M_P^4k^3}\f{e^{4\pi\xi}}{\xi^6}},\\
 \\
 \ds{\langle\hat{h}^\pm(\vec{k}_1,t)\hat{h}^\pm(\vec{k}_2,t)\rangle_{\Md_1}^{\text{quant}}}&=&\ds{(2\pi)^3\delta^3\left(\vec{k}_1+\vec{k}_2\right)\times\left(6.2\times10^{-5}\right)\f{H^4}{M_P^4k^3}\f{e^{2\pi\xi}}{\xi^3}\ln\left(\f{k}{aH}\right)}.
\end{array}
\end{equation}
The parity violation of the instability is evident from the coefficient discrepancy between the amplitudes for the two different helicities.  Furthermore we conclude, as previous researchers have argued \cite{Biagetti:2014asa}, that the one vertex diagram is small compared to the two vertex diagram.  Note it appears that for small $\xi\ll3$ the one vertex diagram dominates over the two vertex diagram, however in that case both diagrams are sub-dominant to the usual tree result and the expressions reported here are not valid.

We want to also emphasize that the U(1) coupled to the pseudoscalar could be a dark unbroken U(1) and the effect on the tensor two point function would still proceed as we have described.

\section{Semiclassical contribution}\label{sec:Classical}

The semiclassical diagrams, due to the presence of an inhomogeneous background, produce a more complicated momentum dependence in the correlation function since there may be net momentum transfer between the vector field and the tensor fluctuations.  However, we will show that the required properties of the classical vector field in order to leave observable imprints may be difficult to satisfy.

The initial conditions of the classical vector field at the onset of inflation are model dependent.  However, there are some theoretical consistency bounds that we may impose.  In order for inflation to proceed, the energy density of the vector field must be less than the potential energy of the inflaton.  Furthermore, we do not want vector backreaction on the pseudoscalar field since that may disrupt its slow rolling \cite{Barnaby:2012xt}.\footnote{Note, it has been previously shown that satifying $\rho_A\ll\dot{\chi}^2/2$ implies that the vector backreaction term may be ignored from the pseudoscalar equation of motion \cite{Barnaby:2012xt} for the quantum fluctuations.  In principle one only needs to require that $\vec{E}\cdot\vec{B}\ll\dot{\chi}^2/2$ to neglect the source term in the $\chi$ equation of motion and $\rho_A<\dot{\phi}^2/2$ to ensure that $-\dot{H}/H^2=\epsilon_\phi$.  Analysis of how the perturbation spectrum is changed for $\dot{\phi}^2/2<\rho_A<\rho_\phi$ is an interesting issue which must be studied carefully.}  Therefore we require
\begin{equation}\label{eq:vec_BR}
\rho_A\ll\f{1}{2}\dot{\chi}^2\ll\epsilon H^2M_P^2.
\end{equation}

The momentum space distribution of the classical vector field energy density should likely be centered around a physical momentum which does not greatly exceed the Hubble scale.  The exact details of how this classical vector configuration is generated prior to the onset of inflation will not be discussed here, as we are only concerned as to whether it could have observable effects.

Recent analysis of the transition to inflation has indicated that there are typically eight or more efolds of inflation which must preceed the scales we observe in the CMB exiting the horizon \cite{Aravind:2016bnx}.  For a transition timescale which is slow, this reduces to approximately three efolds.  However, typical models for the onset of inflation are usually well approximated as rapid transitions.  This implies that the energy density of the vector field is significantly redshifted prior to our cosmologically observable modes exiting the horizon since $\rho\propto a^{-4}$.  However, this will be the case for both the classical and quantum fluctuations of the vector field.  Therefore the relevant comparison is how the classical $F_{\mu\nu}$ in the interaction Hamiltonian compares to the fluctuation of $F_{\mu\nu}$ for physical momenta scales $p\gtrsim 3000H$.\footnote{For the case of a very slow transition to inflation, this becomes $p\gtrsim20H$.}

For quantum fluctuations, the electric field prior to enhancement is non-vanishing and scales as $F_{0i,p}^{\text{fluct}}\propto\sqrt{p}$.  Therefore the classical vector field should be excited in such a way as to have $F_{0i,p}^{\text{classical}}\gg F_{0i,p}^{\text{fluct}}$ for $p\gtrsim3000H$.\footnote{A similar discussion holds for the magnetic field contribution.}  We have been unable to construct such a mechanism for exciting classical vector fields with this property.  Since the total energy density of the vector field is bounded by backreaction considerations \eqref{eq:vec_BR}, the most optmistic case corresponds to only a few modes (each with $p\gtrsim 3000H$) of the classical vector field being excited.  Determining a concrete mechanism for generating such a configuration so that detailed predictions may be made remains an interesting direction for future work.

\section{Conclusion}\label{sec:Conclusion}
A current challenge for inflationary cosmology is to ensure that any future positive detection of a primordial tensor spectrum is properly interpreted.  It has been shown that simply by allowing for a spectator pseudoscalar field the interpretation may be made ambiguous due to amplified vector fields running in quantum loops.  Since this ambiguity is the result of an instability that develops in vector fields coupled to the pseudoscalar through a Chern-Simons term, one may ask if the presence of an initial classical vector field component can effect observable quantities through semiclassical contributions.

In this paper we have discussed the amplification of inhomogeneous vector backgrounds which were generated prior to the onset of inflation, showing that the properties that must be satisfied by the classical vector background in order to have observable imprints may be difficult to satisfy.  Recent studies show that there are typically eight or more efolds of inflation after the onset of inflation before our cosmologically observable modes exit the horizon \cite{Aravind:2016bnx}, causing the scales $p\lesssim3000H$ to be irretrievably pushed outside of our observable horizon.  This means a mechanism for exciting a classical vector background on scales $p\gtrsim3000H$ prior to the onset of inflation must be constructed with an electric field amplitude which dominates over the quantum fluctuation in electric field .  We were unable to construct such a mechanism, though it remains an interesting direction for future studies.

\section*{Acknowledgments}
We would like to thank Lorenzo Sorbo for helpful discussions.  We would also like to thank Ricardo Ferreira and Martin Sloth for a helpful correspondence.  This material is based upon work supported by the National Science Foundation under Grant Number PHY-1316033.

%


\newpage
\begin{thebibliography}{19}        



\bibitem{Baumann:2008aq} 
D.~Baumann {\it et al.}  [CMBPol Study Team Collaboration],
``CMBPol Mission Concept Study: Probing Inflation with CMB Polarization,''
AIP Conf.\ Proc.\  {\bf 1141}, 10 (2009)
\href{http://arxiv.org/abs/0811.3919}{[arXiv:0811.3919 [astro-ph]]}.

\bibitem{Barnaby:2012xt} 
  N.~Barnaby, J.~Moxon, R.~Namba, M.~Peloso, G.~Shiu and P.~Zhou,
  ``Gravity waves and non-Gaussian features from particle production in a sector gravitationally coupled to the inflaton,''
  Phys.\ Rev.\ D {\bf 86}, 103508 (2012)
  \href{http://arxiv.org/abs/1206.6117}{[arXiv:1206.6117 [astro-ph.CO]]}.

\bibitem{Carroll:1989vb} 
  S.~M.~Carroll, G.~B.~Field and R.~Jackiw,
  ``Limits on a Lorentz and Parity Violating Modification of Electrodynamics,''
  Phys.\ Rev.\ D {\bf 41}, 1231 (1990).
  
\bibitem{Maleknejad:2012fw} 
A.~Maleknejad, M.~M.~Sheikh-Jabbari and J.~Soda,
``Gauge Fields and Inflation,''
Phys.\ Rept.\  {\bf 528}, 161 (2013)
\href{http://arxiv.org/abs/1212.2921}{[arXiv:1212.2921 [hep-th]]}.

\bibitem{Anber:2006xt} 
  M.~M.~Anber and L.~Sorbo,
  ``N-flationary magnetic fields,''
  JCAP {\bf 0610}, 018 (2006)
  \href{http://arxiv.org/abs/astro-ph/0606534}{[astro-ph/0606534]}.

\bibitem{Bartolo:2014hwa} 
  N.~Bartolo, S.~Matarrese, M.~Peloso and M.~Shiraishi,
  ``Parity-violating and anisotropic correlations in pseudoscalar inflation,''
  JCAP {\bf 1501}, no. 01, 027 (2015)
  \href{http://arxiv.org/abs/1411.2521}{[arXiv:1411.2521 [astro-ph.CO]]}.

\bibitem{Sorbo:2011rz} 
  L.~Sorbo,
  ``Parity violation in the Cosmic Microwave Background from a pseudoscalar inflaton,''
  JCAP {\bf 1106}, 003 (2011)
  \href{http://arxiv.org/abs/1101.1525}{[arXiv:1101.1525 [astro-ph.CO]]}.
  
\bibitem{GarciaBellido:1995fz} 
  J.~Garcia-Bellido and D.~Wands,
  ``Constraints from inflation on scalar - tensor gravity theories,''
  Phys.\ Rev.\ D {\bf 52}, 6739 (1995)
  \href{http://arxiv.org/abs/gr-qc/9506050}{[gr-qc/9506050]}.
  
\bibitem{Taruya:1997iv} 
  A.~Taruya and Y.~Nambu,
  ``Cosmological perturbation with two scalar fields in reheating after inflation,''
  Phys.\ Lett.\ B {\bf 428}, 37 (1998)
  \href{http://arxiv.org/abs/gr-qc/9709035}{[gr-qc/9709035]}.
  
\bibitem{Finelli:2000ya} 
  F.~Finelli and R.~H.~Brandenberger,
  ``Parametric amplification of metric fluctuations during reheating in two field models,''
  Phys.\ Rev.\ D {\bf 62}, 083502 (2000)
  \href{http://arxiv.org/abs/hep-ph/0003172}{[hep-ph/0003172]}.
  
\bibitem{Gordon:2000hv} 
  C.~Gordon, D.~Wands, B.~A.~Bassett and R.~Maartens,
  ``Adiabatic and entropy perturbations from inflation,''
  Phys.\ Rev.\ D {\bf 63}, 023506 (2001)
  \href{http://arxiv.org/abs/astro-ph/0009131}{[astro-ph/0009131]}.
  
\bibitem{Amendola:2001ni} 
  L.~Amendola, C.~Gordon, D.~Wands and M.~Sasaki,
  ``Correlated perturbations from inflation and the cosmic microwave background,''
  Phys.\ Rev.\ Lett.\  {\bf 88}, 211302 (2002)
  \href{http://arxiv.org/abs/astro-ph/0107089}{[astro-ph/0107089]}.
  
\bibitem{Malik:2008im} 
  K.~A.~Malik and D.~Wands,
  ``Cosmological perturbations,''
  Phys.\ Rept.\  {\bf 475}, 1 (2009)
  \href{http://arxiv.org/abs/0809.4944}{[arXiv:0809.4944 [astro-ph]]}.
  
\bibitem{Ferreira:2014zia} 
  R.~Z.~Ferreira and M.~S.~Sloth,
  ``Universal Constraints on Axions from Inflation,''
  JHEP {\bf 1412}, 139 (2014)
  \href{http://arxiv.org/abs/1409.5799}{[arXiv:1409.5799 [hep-ph]]}.
  
\bibitem{Namba:2015gja} 
  R.~Namba, M.~Peloso, M.~Shiraishi, L.~Sorbo and C.~Unal,
  ``Scale-dependent gravitational waves from a rolling axion,''
  \href{http://arxiv.org/abs/1509.07521}{arXiv:1509.07521 [astro-ph.CO]}.
  
\bibitem{Weinberg:2005vy} 
  S.~Weinberg,
  ``Quantum contributions to cosmological correlations,''
  Phys.\ Rev.\ D {\bf 72}, 043514 (2005)
  \href{http://arxiv.org/abs/hep-th/0506236}{[hep-th/0506236]}.
  
\bibitem{Adshead:2009cb} 
  P.~Adshead, R.~Easther and E.~A.~Lim,
  ``The 'in-in' Formalism and Cosmological Perturbations,''
  Phys.\ Rev.\ D {\bf 80}, 083521 (2009)
  \href{http://arxiv.org/abs/0904.4207}{[arXiv:0904.4207 [hep-th]]}.
  
\bibitem{Weinberg:2010wq} 
  S.~Weinberg,
  ``Ultraviolet Divergences in Cosmological Correlations,''
  Phys.\ Rev.\ D {\bf 83}, 063508 (2011)
  \href{http://arxiv.org/abs/1011.1630}{[arXiv:1011.1630 [hep-th]]}.

\bibitem{Barnaby:2011vw} 
  N.~Barnaby, R.~Namba and M.~Peloso,
  ``Phenomenology of a Pseudo-Scalar Inflaton: Naturally Large Nongaussianity,''
  JCAP {\bf 1104}, 009 (2011)
  \href{http://arxiv.org/abs/1102.4333}{[arXiv:1102.4333 [astro-ph.CO]]}.
  
\bibitem{Anber:2012du} 
  M.~M.~Anber and L.~Sorbo,
  ``Non-Gaussianities and chiral gravitational waves in natural steep inflation,''
  Phys.\ Rev.\ D {\bf 85}, 123537 (2012)
  \href{http://arxiv.org/abs/1203.5849}{[arXiv:1203.5849 [astro-ph.CO]]}.
  
\bibitem{Biagetti:2014asa} 
  M.~Biagetti, E.~Dimastrogiovanni, M.~Fasiello and M.~Peloso,
  ``Gravitational Waves and Scalar Perturbations from Spectator Fields,''
  JCAP {\bf 1504}, no. 04, 011 (2015)
  \href{http://arxiv.org/abs/1411.3029}{[arXiv:1411.3029 [astro-ph.CO]]}.
  
\bibitem{Aravind:2016bnx} 
  A.~Aravind, D.~Lorshbough and S.~Paban,
  ``On primordial equation of state transitions,''
  \href{https://arxiv.org/abs/1604.03516}{arXiv:1604.03516 [astro-ph.CO]}.
\end{thebibliography}
\end{document}